\documentclass[letter]{aa}  

\usepackage{graphicx}
\usepackage{txfonts}

\usepackage{amssymb}
\usepackage{epstopdf}
\usepackage{epsfig}
\usepackage{url}
\usepackage{natbib}
\bibpunct{(}{)}{;}{a}{}{,} 

\begin{document}

   \title{The ALMA-PILS survey: First detections of deuterated formamide and deuterated isocyanic acid in the interstellar medium} 

  \titlerunning {The ALMA-PILS survey: First detections of deuterated formamide and deuterated isocyanic acid}

 \author{A. Coutens \inst{1}  
         \and          J. K. J\o rgensen\inst{2} \and
          M. H. D. van der Wiel\inst{2} \and
          H. S. P. M\"uller\inst{3} \and     
          J. M. Lykke\inst{2} \and
          P. Bjerkeli\inst{2,4} \and   
          T. L. Bourke\inst{5} \and
          H. Calcutt\inst{2} \and
          M. N. Drozdovskaya\inst{6} \and
          C. Favre\inst{7} \and
          E. C. Fayolle\inst{8} \and
          R. T. Garrod\inst{9} \and
          S. K. Jacobsen\inst{2} \and
          N. F. W. Ligterink\inst{6} \and
          K. I. \"Oberg\inst{8} \and
          M. V. Persson\inst{6} \and
          E. F. van Dishoeck\inst{6,10} \and
          S. F. Wampfler\inst{2}
          }

   \institute{Department of Physics and Astronomy, University College London, Gower St., London, WC1E 6BT, UK \\
   \email{a.coutens@ucl.ac.uk}
   \and   
   Centre for Star and Planet Formation, Niels Bohr Institute \& Natural History Museum of Denmark, University of Copenhagen, {\O}ster Voldgade 5-7, DK-1350 Copenhagen K., Denmark
    \and   
    I. Physikalisches Institut, Universit\"at zu K\"oln, Z\"ulpicher Str. 77, 50937 K\"oln, Germany
    \and  
    Department of Earth and Space Sciences, Chalmers University of Technology, Onsala Space Observatory, 439 92 Onsala, Sweden 
    \and
    SKA Organization, Jodrell Bank Observatory, Lower Withington, Macclesfield, Cheshire SK11 9DL, UK
    \and
    Leiden Observatory, Leiden University, PO Box 9513, NL-2300 RA Leiden, the Netherlands
    \and 
    Institut de Plan\'etologie et d'Astrophysique de Grenoble, UMR 5274, UJF-Grenoble 1/CNRS, 38041 Grenoble, France
    \and 
    Harvard-Smithsonian Center for Astrophysics, 60 Garden Street, Cambridge, MA 02138, USA
    \and
     Departments of Chemistry and Astronomy, University of Virginia, Charlottesville, VA 22904, USA
    \and  
    Max-Planck Institut f\"ur Extraterrestrische Physik (MPE), Giessenbachstr. 1, 85748 Garching, Germany
             }

   \date{Received xxx; accepted xxx}

  \abstract{ 
  Formamide (NH$_2$CHO) has previously been detected in several star-forming regions and is thought to be a precursor for different prebiotic molecules. Its formation mechanism is still debated, however.
 Observations of formamide, related species, and their isopotologues may provide useful clues to the chemical pathways leading to their formation. The Protostellar Interferometric Line Survey (PILS) represents an unbiased, high angular resolution and sensitivity spectral survey of the low-mass protostellar binary IRAS~16293--2422 with the Atacama Large Millimeter/submillimeter Array (ALMA). 
 For the first time, we detect  the three singly deuterated forms of NH$_2$CHO (NH$_2$CDO, cis- and trans-NHDCHO), as well as DNCO towards the component B of this binary source. The images reveal that the different isotopologues  are all present in the same region. Based on  observations of the $^{13}$C isotopologues of formamide and a standard $^{12}$C/$^{13}$C ratio, the deuterium fractionation is found to be similar for the three different forms with a value of about 2\%. The DNCO/HNCO ratio is also comparable to the D/H ratio of formamide ($\sim$1\%). These results are in agreement with the hypothesis that NH$_2$CHO and HNCO are chemically related through grain-surface formation.}    

   \maketitle
%

\section{Introduction}

Formamide (NH$_2$CHO), also known as methanamide, contains the amide bond (--N--C(=O)--), which plays an important role in the synthesis of proteins. This molecule is a precursor for potential compounds of genetic and metabolic interest \citep{Saladino2012}. Interestingly, it is present in various astrophysical environments: high-mass star-forming regions \citep[e.g.][]{Bisschop2007,Adande2013}, low-mass protostars \citep{Kahane2013,Lopez2015}, shocked regions \citep{Yamaguchi2012,Mendoza2014}, a translucent cloud \citep{Corby2015}, comets \citep{Bockelee2000,Biver2014,Goesmann2015}, and even an extragalactic source  \citep{Muller2013}. 

The formation of formamide is still not clearly understood: several routes have been proposed, both in the gas phase and on the grain surfaces.
In the gas phase, many ion-molecule reactions have been ruled out as not sufficiently efficient owning to endothermicity or high-energy barriers (see e.g. \citealt{Redondo2014a,Redondo2014b}). A neutral-neutral reaction between H$_2$CO and NH$_2$ was, however, shown to be barrierless and could account for the abundance of formamide in some sources \citep{Barone2015}. On the grain surface, it has been suggested that formamide forms through the reaction between HCO and NH$_2$ \citep{Jones2011,Garrod2013} and/or hydrogenation of isocyanic acid, HNCO. In particular, the latter suggestion is supported by a strong correlation between the HNCO and NH$_2$CHO abundances in different sources \citep{Bisschop2007,Mendoza2014,Lopez2015}. However, an experiment based on the H-bombardment of HNCO at low temperature has recently shown that this reaction is not efficient in cold environments \citep{Noble2015}. Instead, other pathways to HNCO and NH$_2$CHO on grains have been
 suggested, either with or without UV or ion bombardment (see e.g. \citealt{Kanuchova2016} and references therein). 
 
Measurements of isotopic fractionation may help to constrain formation pathways of molecules since isotopic fractionation (especially deuteration) is sensitive to physical conditions such as density and temperature. Until recently, the study of deuteration in solar-type protostars was mainly limited to relatively small and abundant molecules, such as H$_2$O, HCO$^+$, HCN, H$_2$CO, and CH$_3$OH. Even though the deuterium fractionation is known to be enhanced in low-mass protostars \citep[see e.g.][]{Ceccarelli2007}, measurements of lines of deuterated complex organic molecules (COMs) still require high-sensitivity observations. So far, only deuterated methyl formate and dimethyl ether have been detected towards the low-mass protostar IRAS~16293--2422 (hereafter IRAS16293) by \citet{Demyk2010} and \citet{Richard2013}. With the Atacama Large Millimeter/submillimeter Array (ALMA), it is now possible to search for the isotopologues of complex and less abundant species. In this Letter, we report the first detection of the three singly deuterated forms of formamide, as well as DNCO towards IRAS16293. These observations mark the first detections of those isotopologues in the interstellar medium.

\section{Observations}
\label{sect_obs}

An ALMA unbiased spectral survey of the binary protostar IRAS16293 was recently carried out in the framework of the Protostellar Interferometric Line Survey\footnote{\url{http://youngstars.nbi.dk/PILS/}} (PILS;  \citealt{Jorgensen2016}). The observations were centered on a position at equal distance between the sources A and B that are separated by $\sim$5$\arcsec$. A full description of the survey and the data reduction can be found in \citet{Jorgensen2016}. For this work, we use the part of the large spectral survey obtained in Band 7 between 329.15 GHz and 362.90 GHz, both with the 12m array and the Atacama Compact Array (ACA). 
The spectral resolution of these observations is 0.244 MHz (i.e. $\sim$0.2 km\,s$^{-1}$). 
After combining the 12m and ACA data, the final spectral line datacubes show a sensitivity that is better than 5 mJy beam$^{-1}$ km\,s$^{-1}$. The beam sizes range between 0.4$\arcsec$ and 0.7$\arcsec$. Additional observations in Bands 3 and 6 cover narrow spectral ranges and, consequently, a very limited number of transitions of formamide isotopologues. After the analysis of Band 7, we checked that the results are consistent with these lower frequency observations.

\section{Analysis and results}
\label{sect_analysis}

To search for the isotopologues of formamide, we use the spectrum extracted at the same position as in \citet{Lykke2016}, i.e. a position offset by $\sim$0.5$\arcsec$ from the continuum peak of source B in the south west direction ($\alpha_{\rm J2000}$=16$^{\rm h}$32$^{\rm m}$22$\fs$58, $\delta_{\rm J2000}$=-24$\degr$28$\arcmin$32.8$\arcsec$). Although the lines are brighter at the position of the continuum peak, the presence of both absorption and emission makes analysis difficult. At the selected position, most of the lines present Gaussian profiles and are relatively bright compared to other positions. In source A, the lines are quite broad, leading to significant line confusion that prevents the search for isotopologues of complex species \citep[e.g.][]{Jorgensen2012}. This Letter is therefore focused on source B only.

\begin{figure}
\centering
\includegraphics[width=0.85\columnwidth]{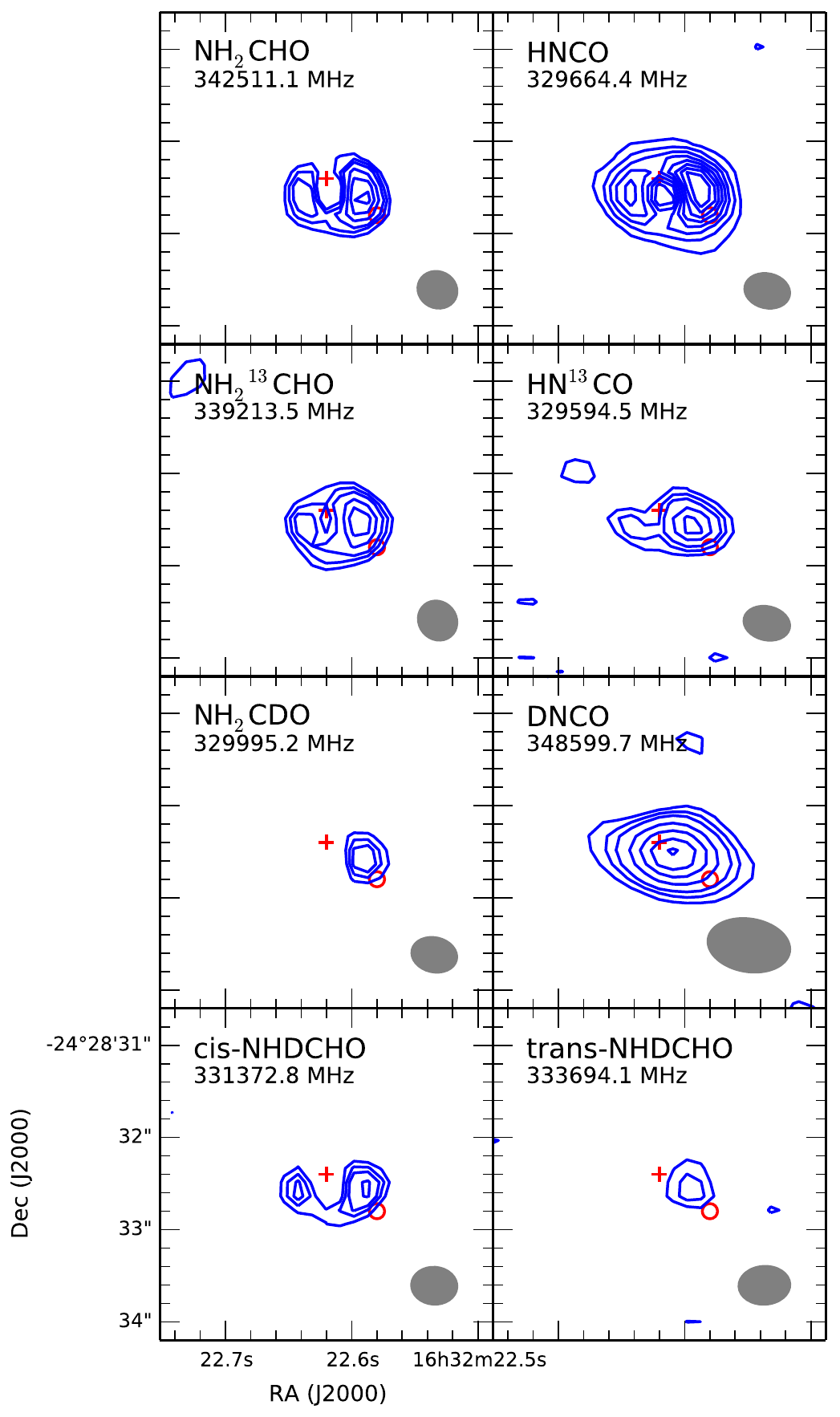}
\caption{Integrated intensity maps of NH$_2$CHO, HNCO, and their isotopologues towards source B. The position of the continuum peak of source B is indicated with a red cross, while the position of where the spectrum was extracted is shown with a red circle. The beam sizes are shown in gray in the bottom right corner of each panel. The contour levels start for the main isotopologue of HNCO at 0.05 Jy\,km\,s$^{-1}$ with a step of 0.05 Jy\,km\,s$^{-1}$. For the other species, the levels are 0.02, 0.03, 0.04, 0.06, 0.08, 0.1, and 0.12 Jy\,km\,s$^{-1}$. }
\label{fig:maps}
\end{figure}

Using the CASSIS\footnote{\url{http://cassis.irap.omp.eu/}} software, we identify several unblended lines that can be assigned to the three singly deuterated forms of NH$_2$CHO and to NH$_2$$^{13}$CHO, DNCO, and HN$^{13}$CO (see Table~\ref{results_model}). These mark the first detections of NH$_2$CDO, cis-NHDCHO, trans-NHDCHO, and DNCO in the interstellar medium. The list of unblended lines can be found in the Appendix. Maps of the integrated line emission from representative lines from the different isotopologues towards source B are shown in Figure \ref{fig:maps}. The emission of the different lines clearly arise from a similar compact region in the vicinity of IRAS16293B. A hole is observed in the maps owing to the absorptions that are produced against the strong continuum at the continuum peak position. For DNCO, the larger beam size for the observations of this transition masks the absorption. The spatial variations that are observed among the different species are probably due to different line excitation or line brightness. In particular, HNCO seems to be slightly more extended than NH$_2$CHO, but this is most likely due to the fact that the HNCO lines are particularly bright compared to the HNCO and formamide isotopologues.

\begin{figure}
\includegraphics[width=0.90\columnwidth]{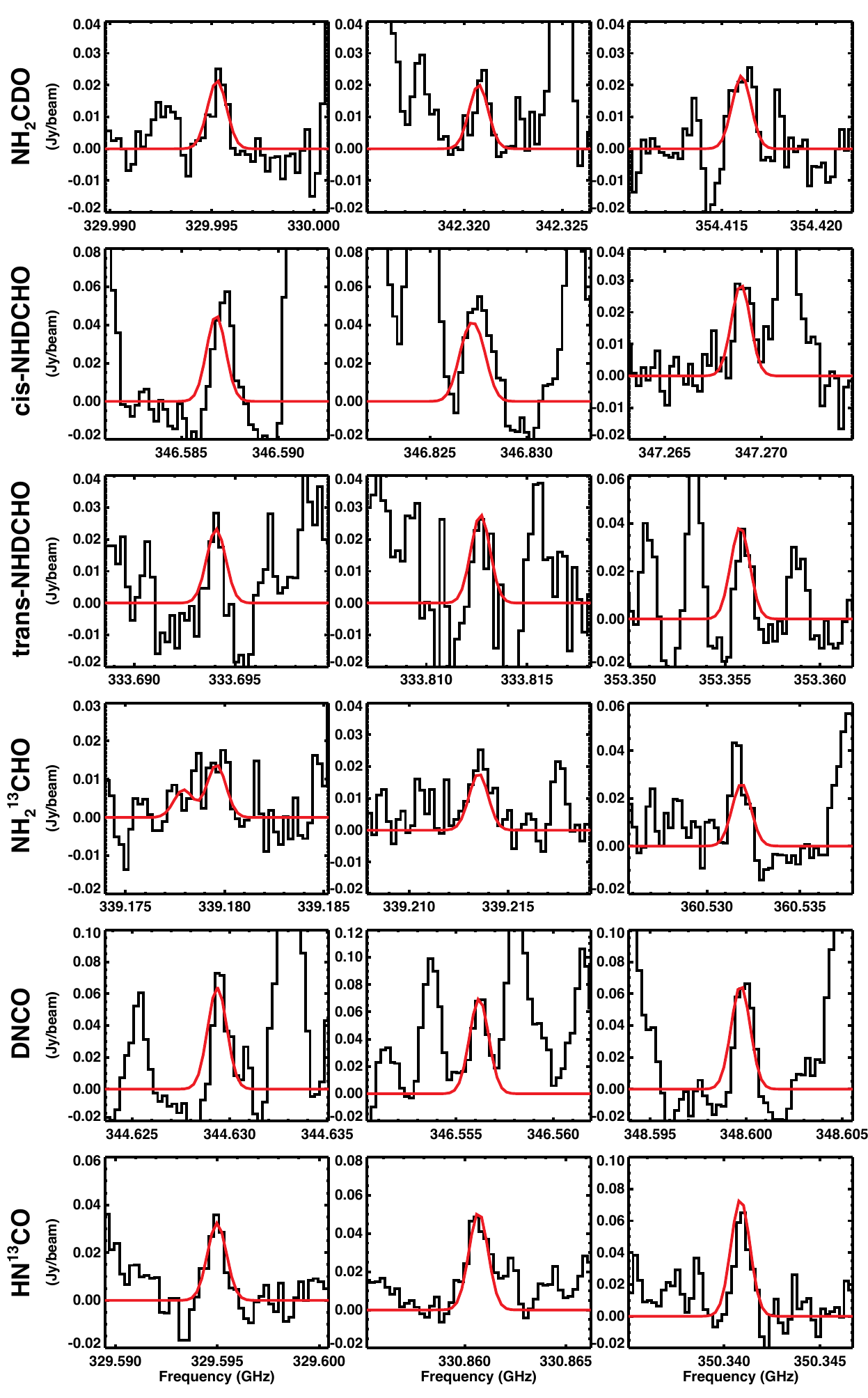}
\caption{{\it Black}: Detected lines of NH$_2$CDO, cis-NHDCHO, trans-NHDCHO, NH$_2$$^{13}$CHO, DNCO, and HN$^{13}$CO. {\it Red}: Best-fit model.}
\label{fig:all_model}
\end{figure}

To constrain the excitation temperatures and column densities of the different species, we produce a grid of synthetic spectra, assuming local thermodynamical equilibrium (LTE). We predict the spectra for different excitation temperatures between 100 and 300 K with a step of 25 K and for different column densities between 1\,$\times$\,10$^{13}$ and 1\,$\times$\,10$^{17}$ cm$^{-2}$. First, the column density is roughly estimated using relatively large steps, then refined using smaller steps around the best-fit solution. We determine the best-fit model using a $\chi$$^2$ method, comparing the observed and synthetic spectra at $\pm$0.5 MHz around the rest frequency of the predicted emission lines. We carefully check that the best-fit model does not predict any lines that are not observed in the spectra.
For the deuterated forms, the models are in agreement with the observations for excitation temperatures between 100 and 300 K. However, for NH$_2$$^{13}$CHO and HN$^{13}$CO, a model with a high excitation temperature better reproduces the observed emission than a model with a low excitation temperature (see Figs \ref{fig:NH213CHO} and \ref{fig:HN13CO}). An excitation temperature of 300~K was consequently adopted for the analysis of the different isotopologues.  
This excitation temperature is similar to that derived for glycolaldehyde and ethylene glycol \citep[][submitted]{Jorgensen2012}, but higher than what is found for acetaldehyde, ethylene oxide, and propanal ($\sim$125\,K, \citealt{Lykke2016}). The derived column densities, assuming a linewidth of 1 km\,s$^{-1}$ and a source size of 0.5$\arcsec$ \citep{Jorgensen2016,Lykke2016}, are summarized in Table \ref{results_model}. The uncertainties on the column densities are all estimated to be within a factor of 2 (including the uncertainty on both the excitation temperature and the baseline subtraction). The upper limits are estimated visually by a comparison of the synthetic spectra with the observations on the entire spectral range. Figure \ref{fig:all_model} shows three lines of each isotopologue with the best-fit model. The models for all the lines are shown in Appendix B.

\begin{table}
\caption{Number of lines used in the analysis of the isotopologues of NH$_2$CHO and HNCO and column densities derived for $T_{\rm ex}$\,=\,300\,K and a source size of 0.5$\arcsec$.}
\label{results_model}
\centering
\begin{tabular}{l c c c }
\hline
Species & $\#$ of lines & $E_{\rm up}$ (K) & $N$ (cm$^{-2}$) \\
\hline
NH$_2$CDO     & 12 & 146 -- 366 & 2.1\,$\times$\,10$^{14}$   \\ 
cis-NHDCHO     & 11 & 146 -- 307 & 2.1\,$\times$\,10$^{14}$   \\ 
trans-NHDCHO & 11 & 151 -- 332 & 1.8\,$\times$\,10$^{14}$    \\ 
NH$_2$$^{13}$CHO & 10 & 152 -- 428 & 1.5\,$\times$\,10$^{14}$ \\ 
$^{15}$NH$_2$CHO  & -- & -- & $\leq$ 1.0\,$\times$\,10$^{14}$ $^{(a)}$ \\
NH$_2$CH$^{18}$O & -- & -- & $\leq$ 0.8\,$\times$\,10$^{14}$ $^{(a)}$ \\ %
DNCO & 4 & 150 -- 751 & 3.0\,$\times$\,10$^{14}$ \\ %
HN$^{13}$CO & 8 & 127 -- 532 & 4.0\,$\times$\,10$^{14}$ \\ %
H$^{15}$NCO & -- & -- & $\leq$ 2.0\,$\times$\,10$^{14}$ $^{(a)}$\\
HNC$^{18}$O & --  & --  & $\leq$ 1.5\,$\times$\,10$^{14}$ $^{(a)}$ \\ 
\hline \hline
\end{tabular}
\vspace{-0.2cm}
\tablefoot{
\tablefoottext{a} {3$\sigma$ upper limit.}}
\end{table}%

The column densities of NH$_2$$^{13}$CHO, and HN$^{13}$CO are estimated to be 1.5\,$\times$\,10$^{14}$ cm$^{-2}$ and 4\,$\times$\,10$^{14}$ cm$^{-2}$, respectively. Assuming a $^{12}$C/$^{13}$C ratio of 68 \citep{Milam2005}, the column densities for the main isotopologues of formamide and isocyanic acid are predicted to be 1\,$\times$\,10$^{16}$ cm$^{-2}$ and 3\,$\times$\,10$^{16}$ cm$^{-2}$. With these column densities, several NH$_2$CHO lines and all of the HNCO lines are overproduced, indicating that they are optically thick. The model of formamide is, however, in agreement with the few lines with the lowest opacities (see Figs. \ref{fig:HCONH2_v0} and \ref{fig:HCONH2_v12}). NH$_2$CH$^{18}$O has also been searched for, but is not detected with a 3$\sigma$ upper limit of 8\,$\times$\,10$^{13}$ cm$^{-2}$. The non-detection of this isotopologue is consistent with the $^{16}$O/$^{18}$O ratio of 560 in the interstellar medium \citep[][]{Wilson1999}, which gives $N$(NH$_2$CH$^{18}$O) = 2\,$\times$\,10$^{13}$ cm$^{-2}$. Similarly, HNC$^{18}$O is not detected either with a 3$\sigma$ upper limit of 1.5\,$\times$\,10$^{14}$ cm$^{-2}$, which is consistent with its expected column density of 5\,$\times$\,10$^{13}$ cm$^{-2}$.

Using the column densities derived for the $^{13}$C isotopologues and a standard $^{12}$C/$^{13}$C ratio, the deuterium fractionation in NH$_2$CHO is about 2\% for the three deuterated forms and the DNCO/HNCO ratio is similar ($\sim$1\%). If the $^{12}$C/$^{13}$C ratio is lower ($\sim$30) as reported for glycolaldehyde by \citet{Jorgensen2016}, the D/H ratios of formamide and isocyanic acid would be about 4-5\% and 2-3\%, respectively.

We also search for the $^{15}$N isotopologues of formamide and isocyanic acid. A couple of transitions could tentatively be assigned to $^{15}$NH$_2$CHO, but these lines are close to the noise level and possibly blended with other species. For H$^{15}$NCO, the uncertainties on the frequencies of some of the transitions are rather large, preventing any firm detection. Based on a standard $^{12}$C/$^{13}$C ratio, lower limits of 100 and 138 are obtained for the $^{14}$N/$^{15}$N ratios of formamide and HNCO respectively.

\section{Discussion and conclusion}
\label{sect_discussion}

Our derived ratio in IRAS16293 for HNCO/NH$_2$CHO, $\sim$3, is consistent with the ratios found  in warm sources in previous studies \citep{Bisschop2007,Mendoza2014,Lopez2015}. Thanks to our interferometric observations, we also confirm that these two species are spatially correlated. The deuterium fractionation ratios of these two molecules are also similar, reinforcing the hypothesis that they are chemically related.
We discuss here possible scenarios for the formation of these species in the warm inner regions of protostars. 

Assuming that the deuteration of formaldehyde in the region probed by the ALMA observations of formamide is similar to the value derived with single-dish observations ($\sim$15\%, \citealt{Loinard2000}), we can discuss the possibility for the gas-phase formation mechanism proposed by \citet{Barone2015}, H$_2$CO + NH$_2$ $\rightarrow$ NH$_2$CHO + H.
According to this reaction, the deuterated form NHDCHO would result from the reaction between NHD and H$_2$CO, while NH$_2$CDO would form from NH$_2$ and HDCO. 
We would consequently expect a higher deuteration for NH$_2$CDO, compared to the observations, unless the reaction between NH$_2$ and HDCO leads more efficiently to NH$_2$CHO and D, compared to NH$_2$CDO and H. Theoretical or experimental studies of the branching ratios of these reactions would be needed to rule out this scenario. The  determination of the HDCO/H$_2$CO ratio from the PILS survey is also necessary. Nevertheless, it should be noted that, so far, there is no proposed scenario in the gas phase that could explain the correlation with HNCO.

Although it was recently shown that NH$_2$CHO does not form by
 hydrogenation of HNCO on grain surfaces \citep{Noble2015}, several
 other proposed mechanisms exist in the literature. Both species can
 be formed through barrierless reactions in ices through NH + CO
 $\to$ HNCO and NH$_2$ + H$_2$CO $\to$ NH$_2$CHO + H, as
 demonstrated experimentally \citep{Fedoseev2015,
 Fedoseev2016}. Alternatively, both species are formed through ion bombardment of H$_2$O:CH$_4$:N$_2$ mixtures \citep{Kanuchova2016} or
 UV irradiation of CO:NH$_3$:CH$_3$OH and/or HNCO mixtures \citep[e.g.][]{Demyk1998,Raunier2004,Jones2011,Henderson2015}.  Quantitative gas-grain modeling under conditions
 representative of IRAS16293 are needed to assess which of these
 grain surface routes dominates.
 
Ultimately, the HNCO and NH$_2$CHO deuterium fractionation level and pattern may also hold a clue to their formation routes. A particularly
 interesting result is that the three singly deuterated forms of formamide are found with similar abundances in IRAS16293. 
Contrary to the -CH functional group that is not affected by hydrogen isotope exchanges, the hydroxyl (-OH) and amine (-NH) groups are expected to establish hydrogen bonds and  equilibrate with water \citep{Faure2015}. This mechanism was proposed to explain the different CH$_3$OD/CH$_3$OH ($\sim$1.8\%) and CH$_2$DOH/CH$_3$OH ($\sim$37\%) ratios derived in IRAS16293 \citep{Parise2006}, as the water deuterium fractionation of water in the upper layers of the grain mantles, where complex organic molecules form, is about a few percent \citep{Coutens2012,Coutens2013,Furuya2016}. We do not see these types of differences for formamide, for which all forms show a deuterium fractionation similar to the CH$_3$OD/CH$_3$OH ratio and water. The deuterium fractionation of methanol from the PILS data needs to be investigated to know if the different deuterium fractionation ratios of the -CH and -OH groups are also observed at small scales.

In conclusion, in this Letter we present the first detection of the three singly deuterated forms of formamide and DNCO. The similar deuteration of these species and their similar spatial distributions favors the formation of these two species on grain surfaces. Further studies are, however, needed to rule out gas phase routes. 
These detections illustrate the strength of ALMA, and large spectral surveys such as PILS in particular, for the detections of deuterated complex molecules. Determinations of the deuterium fractionation for more complex molecules will help to constrain their formation pathways. The search for deuterated formamide in more sources is needed to reveal how variable the deuteration of formamide is, and if the similarity of the abundances of the three deuterated forms is common.

\begin{acknowledgements} 
The authors thank Gleb Fedoseev and Harold Linnartz for fruitful discussions. 
This paper makes use of the following ALMA data: ADS/JAO.ALMA\#2013.1.00278.S. ALMA is a partnership of ESO (representing its member states), NSF (USA) and NINS (Japan), together with NRC (Canada) and NSC and ASIAA (Taiwan), in cooperation with the Republic of Chile. The Joint ALMA Observatory is operated by ESO, AUI/NRAO and NAOJ.
The work of AC was funded by an STFC grant. AC thanks the COST action CM1401 Our Astrochemical History for additional financial support.
The group of JKJ acknowledges support from a Lundbeck Foundation Group Leader Fellowship, as well as the European Research Council (ERC) under the European Union's Horizon 2020 research and innovation programme (grant agreement No 646908) through ERC Consolidator Grant S4F. Research at the Centre for Star and Planet Formation is funded by the Danish National Research Foundation. The group of EvD acknowledges A-ERC grant 291141 CHEMPLAN.
\end{acknowledgements}

\bibliographystyle{aa}
\bibliography{biblio_NH2CHO}

\begin{thebibliography}{54}
\expandafter\ifx\csname natexlab\endcsname\relax\def\natexlab#1{#1}\fi

\bibitem[{{Adande} {et~al.}(2013){Adande}, {Woolf}, \& {Ziurys}}]{Adande2013}
{Adande}, G.~R., {Woolf}, N.~J., \& {Ziurys}, L.~M. 2013, Astrobiology, 13, 439

\bibitem[{{Barone} {et~al.}(2015){Barone}, {Latouche}, {Skouteris}, {Vazart},
  {Balucani}, {Ceccarelli}, \& {Lefloch}}]{Barone2015}
{Barone}, V., {Latouche}, C., {Skouteris}, D., {et~al.} 2015, \mnras, 453, L31

\bibitem[{{Bisschop} {et~al.}(2007){Bisschop}, {J{\o}rgensen}, {van Dishoeck},
  \& {de Wachter}}]{Bisschop2007}
{Bisschop}, S.~E., {J{\o}rgensen}, J.~K., {van Dishoeck}, E.~F., \& {de
  Wachter}, E.~B.~M. 2007, \aap, 465, 913

\bibitem[{{Biver} {et~al.}(2014){Biver}, {Bockel{\'e}e-Morvan}, {Debout},
  {Crovisier}, {Boissier}, {Lis}, {Dello Russo}, {Moreno}, {Colom}, {Paubert},
  {Vervack}, \& {Weaver}}]{Biver2014}
{Biver}, N., {Bockel{\'e}e-Morvan}, D., {Debout}, V., {et~al.} 2014, \aap, 566,
  L5

\bibitem[{{Blanco} {et~al.}(2006){Blanco}, {Lopez}, {Lessari}, \&
  {Alonso}}]{Blanco2006}
{Blanco}, S., {Lopez}, J.~C., {Lessari}, A., \& {Alonso}, J.~L. 2006, J. Am.
  Chem. Soc., 128, 12111

\bibitem[{{Bockel{\'e}e-Morvan} {et~al.}(2000){Bockel{\'e}e-Morvan}, {Lis},
  {Wink}, {Despois}, {Crovisier}, {Bachiller}, {Benford}, {Biver}, {Colom},
  {Davies}, {G{\'e}rard}, {Germain}, {Houde}, {Mehringer}, {Moreno}, {Paubert},
  {Phillips}, \& {Rauer}}]{Bockelee2000}
{Bockel{\'e}e-Morvan}, D., {Lis}, D.~C., {Wink}, J.~E., {et~al.} 2000, \aap,
  353, 1101

\bibitem[{{Ceccarelli} {et~al.}(2007){Ceccarelli}, {Caselli}, {Herbst},
  {Tielens}, \& {Caux}}]{Ceccarelli2007}
{Ceccarelli}, C., {Caselli}, P., {Herbst}, E., {Tielens}, A.~G.~G.~M., \&
  {Caux}, E. 2007, Protostars and Planets V, 47

\bibitem[{{Corby} {et~al.}(2015){Corby}, {Jones}, {Cunningham}, {Menten},
  {Belloche}, {Schwab}, {Walsh}, {Balnozan}, {Bronfman}, {Lo}, \&
  {Remijan}}]{Corby2015}
{Corby}, J.~F., {Jones}, P.~A., {Cunningham}, M.~R., {et~al.} 2015, \mnras,
  452, 3969

\bibitem[{{Coutens} {et~al.}(2012){Coutens}, {Vastel}, {Caux}, {Ceccarelli},
  {Bottinelli}, {Wiesenfeld}, {Faure}, {Scribano}, \& {Kahane}}]{Coutens2012}
{Coutens}, A., {Vastel}, C., {Caux}, E., {et~al.} 2012, \aap, 539, A132

\bibitem[{{Coutens} {et~al.}(2013){Coutens}, {Vastel}, {Cazaux}, {Bottinelli},
  {Caux}, {Ceccarelli}, {Demyk}, {Taquet}, \& {Wakelam}}]{Coutens2013}
{Coutens}, A., {Vastel}, C., {Cazaux}, S., {et~al.} 2013, \aap, 553, A75

\bibitem[{{Demyk} {et~al.}(2010){Demyk}, {Bottinelli}, {Caux}, {Vastel},
  {Ceccarelli}, {Kahane}, \& {Castets}}]{Demyk2010}
{Demyk}, K., {Bottinelli}, S., {Caux}, E., {et~al.} 2010, \aap, 517, A17

\bibitem[{{Demyk} {et~al.}(1998){Demyk}, {Dartois}, {D'Hendecourt}, {Jourdain
  de Muizon}, {Heras}, \& {Breitfellner}}]{Demyk1998}
{Demyk}, K., {Dartois}, E., {D'Hendecourt}, L., {et~al.} 1998, \aap, 339, 553

\bibitem[{{Faure} {et~al.}(2015){Faure}, {Faure}, {Theul{\'e}}, {Quirico}, \&
  {Schmitt}}]{Faure2015}
{Faure}, A., {Faure}, M., {Theul{\'e}}, P., {Quirico}, E., \& {Schmitt}, B.
  2015, \aap, 584, A98

\bibitem[{{Fedoseev} {et~al.}(2016){Fedoseev}, {Chuang}, {van Dishoeck},
  {Ioppolo}, \& {Linnartz}}]{Fedoseev2016}
{Fedoseev}, G., {Chuang}, K.-J., {van Dishoeck}, E.~F., {Ioppolo}, S., \&
  {Linnartz}, H. 2016, MNRAS in press

\bibitem[{{Fedoseev} {et~al.}(2015){Fedoseev}, {Ioppolo}, {Zhao}, {Lamberts},
  \& {Linnartz}}]{Fedoseev2015}
{Fedoseev}, G., {Ioppolo}, S., {Zhao}, D., {Lamberts}, T., \& {Linnartz}, H.
  2015, \mnras, 446, 439

\bibitem[{{Furuya} {et~al.}(2016){Furuya}, {van Dishoeck}, \&
  {Aikawa}}]{Furuya2016}
{Furuya}, K., {van Dishoeck}, E.~F., \& {Aikawa}, Y. 2016, \aap, 586, A127

\bibitem[{{Gardner} {et~al.}(1980){Gardner}, {Godfrey}, \&
  {Williams}}]{Gardner1980}
{Gardner}, F.~F., {Godfrey}, P.~D., \& {Williams}, D.~R. 1980, \mnras, 193, 713

\bibitem[{{Garrod}(2013)}]{Garrod2013}
{Garrod}, R.~T. 2013, \apj, 765, 60

\bibitem[{{Goesmann} {et~al.}(2015){Goesmann}, {Rosenbauer}, {Bredeh{\"o}ft},
  {Cabane}, {Ehrenfreund}, {Gautier}, {Giri}, {Kr{\"u}ger}, {Le Roy},
  {MacDermott}, {McKenna-Lawlor}, {Meierhenrich}, {Caro}, {Raulin}, {Roll},
  {Steele}, {Steininger}, {Sternberg}, {Szopa}, {Thiemann}, \&
  {Ulamec}}]{Goesmann2015}
{Goesmann}, F., {Rosenbauer}, H., {Bredeh{\"o}ft}, J.~H., {et~al.} 2015,
  Science, 349, 020689

\bibitem[{{Henderson} \& {Gudipati}(2015)}]{Henderson2015}
{Henderson}, B.~L. \& {Gudipati}, M.~S. 2015, \apj, 800, 66

\bibitem[{{Hirota} {et~al.}(1974){Hirota}, {Sugisaki}, {Nielsen}, \&
  {S{\o}rensen}}]{Hirota1974}
{Hirota}, E., {Sugisaki}, R., {Nielsen}, C.~J., \& {S{\o}rensen}, G.~O. 1974,
  Journal of Molecular Spectroscopy, 49, 251

\bibitem[{{Hocking} {et~al.}(1975){Hocking}, {Gerry}, \&
  {Winnewisser}}]{Hocking1975}
{Hocking}, W.~H., {Gerry}, M.~C.~L., \& {Winnewisser}, G. 1975, Canadian
  Journal of Physics, 53, 1869

\bibitem[{{Jones} {et~al.}(2011){Jones}, {Bennett}, \& {Kaiser}}]{Jones2011}
{Jones}, B.~M., {Bennett}, C.~J., \& {Kaiser}, R.~I. 2011, \apj, 734, 78

\bibitem[{{J{\o}rgensen} {et~al.}(2012){J{\o}rgensen}, {Favre}, {Bisschop},
  {Bourke}, {van Dishoeck}, \& {Schmalzl}}]{Jorgensen2012}
{J{\o}rgensen}, J.~K., {Favre}, C., {Bisschop}, S.~E., {et~al.} 2012, \apjl,
  757, L4

\bibitem[{{J{\o}rgensen} {et~al.}(submitted){J{\o}rgensen}, {van der Wiel},
  {Coutens}, {Lykke}, {M\"uller}, {van Dishoeck}, \& {xxx}}]{Jorgensen2016}
{J{\o}rgensen}, J.~K., {van der Wiel}, M.~H.~D., {Coutens}, A., {et~al.}
  submitted, \aap

\bibitem[{{Kahane} {et~al.}(2013){Kahane}, {Ceccarelli}, {Faure}, \&
  {Caux}}]{Kahane2013}
{Kahane}, C., {Ceccarelli}, C., {Faure}, A., \& {Caux}, E. 2013, \apjl, 763,
  L38

\bibitem[{{Ka{\v n}uchov{\'a}} {et~al.}(2016){Ka{\v n}uchov{\'a}}, {Urso},
  {Baratta}, {Brucato}, {Palumbo}, \& {Strazzulla}}]{Kanuchova2016}
{Ka{\v n}uchov{\'a}}, Z., {Urso}, R.~G., {Baratta}, G.~A., {et~al.} 2016, \aap,
  585, A155

\bibitem[{{Kryvda} {et~al.}(2009){Kryvda}, {Gerasimov}, {Dyubko}, {Alekseev},
  \& {Motiyenko}}]{Kryvda2009}
{Kryvda}, A.~V., {Gerasimov}, V.~G., {Dyubko}, S.~F., {Alekseev}, E.~A., \&
  {Motiyenko}, R.~A. 2009, Journal of Molecular Spectroscopy, 254, 28

\bibitem[{{Kukolich} \& {Nelson}(1971)}]{Kukolich1971}
{Kukolich}, S.~G. \& {Nelson}, A.~C. 1971, Chemical Physics Letters, 11, 383

\bibitem[{{Kurland} \& {Bright Wilson}(1957)}]{Kurland1957}
{Kurland}, R.~J. \& {Bright Wilson}, Jr., E. 1957, \jcp, 27, 585

\bibitem[{{Kutsenko} {et~al.}(2013){Kutsenko}, {Motiyenko}, {Margul{\`e}s}, \&
  {Guillemin}}]{Kutsenko2013}
{Kutsenko}, A.~S., {Motiyenko}, R.~A., {Margul{\`e}s}, L., \& {Guillemin},
  J.-C. 2013, \aap, 549, A128

\bibitem[{{Lapinov} {et~al.}(2007){Lapinov}, {Golubiatnikov}, {Markov}, \&
  {Guarnieri}}]{Lapinov2007}
{Lapinov}, A.~V., {Golubiatnikov}, G.~Y., {Markov}, V.~N., \& {Guarnieri}, A.
  2007, Astronomy Letters, 33, 121

\bibitem[{{Loinard} {et~al.}(2000){Loinard}, {Castets}, {Ceccarelli},
  {Tielens}, {Faure}, {Caux}, \& {Duvert}}]{Loinard2000}
{Loinard}, L., {Castets}, A., {Ceccarelli}, C., {et~al.} 2000, \aap, 359, 1169

\bibitem[{{L{\'o}pez-Sepulcre} {et~al.}(2015){L{\'o}pez-Sepulcre}, {Jaber},
  {Mendoza}, {Lefloch}, {Ceccarelli}, {Vastel}, {Bachiller}, {Cernicharo},
  {Codella}, {Kahane}, {Kama}, \& {Tafalla}}]{Lopez2015}
{L{\'o}pez-Sepulcre}, A., {Jaber}, A.~A., {Mendoza}, E., {et~al.} 2015, \mnras,
  449, 2438

\bibitem[{{Lykke} {et~al.}(to be submitted){Lykke}, {Coutens}, {J{\o}rgensen},
  {xxx}, {xxx}, \& {xxx}}]{Lykke2016}
{Lykke}, J.~M., {Coutens}, A., {J{\o}rgensen}, J.~K., {et~al.} to be submitted

\bibitem[{{Mendoza} {et~al.}(2014){Mendoza}, {Lefloch}, {L{\'o}pez-Sepulcre},
  {Ceccarelli}, {Codella}, {Boechat-Roberty}, \& {Bachiller}}]{Mendoza2014}
{Mendoza}, E., {Lefloch}, B., {L{\'o}pez-Sepulcre}, A., {et~al.} 2014, \mnras,
  445, 151

\bibitem[{{Milam} {et~al.}(2005){Milam}, {Savage}, {Brewster}, {Ziurys}, \&
  {Wyckoff}}]{Milam2005}
{Milam}, S.~N., {Savage}, C., {Brewster}, M.~A., {Ziurys}, L.~M., \& {Wyckoff},
  S. 2005, \apj, 634, 1126

\bibitem[{{Moskienko} \& {Dyubko}(1991)}]{Moskienko1991}
{Moskienko}, E.~M. \& {Dyubko}, S.~F. 1991, Radiophysics and Quantum
  Electronics, 34, 181

\bibitem[{{Motiyenko} {et~al.}(2012){Motiyenko}, {Tercero}, {Cernicharo}, \&
  {Margul{\`e}s}}]{Motiyenko2012}
{Motiyenko}, R.~A., {Tercero}, B., {Cernicharo}, J., \& {Margul{\`e}s}, L.
  2012, \aap, 548, A71

\bibitem[{{M{\"u}ller} {et~al.}(2005){M{\"u}ller}, {Schl{\"o}der}, {Stutzki},
  \& {Winnewisser}}]{Muller2005}
{M{\"u}ller}, H.~S.~P., {Schl{\"o}der}, F., {Stutzki}, J., \& {Winnewisser}, G.
  2005, Journal of Molecular Structure, 742, 215

\bibitem[{{M{\"u}ller} {et~al.}(2001){M{\"u}ller}, {Thorwirth}, {Roth}, \&
  {Winnewisser}}]{Muller2001}
{M{\"u}ller}, H.~S.~P., {Thorwirth}, S., {Roth}, D.~A., \& {Winnewisser}, G.
  2001, \aap, 370, L49

\bibitem[{{Muller} {et~al.}(2013){Muller}, {Beelen}, {Black}, {Curran},
  {Horellou}, {Aalto}, {Combes}, {Gu{\'e}lin}, \& {Henkel}}]{Muller2013}
{Muller}, S., {Beelen}, A., {Black}, J.~H., {et~al.} 2013, \aap, 551, A109

\bibitem[{{Niedenhoff} {et~al.}(1995){Niedenhoff}, {Yamada}, {Belov}, \&
  {Winnewisser}}]{Niedenhoff1995}
{Niedenhoff}, M., {Yamada}, K.~M.~T., {Belov}, S.~P., \& {Winnewisser}, G.
  1995, Journal of Molecular Spectroscopy, 174, 151

\bibitem[{{Noble} {et~al.}(2015){Noble}, {Theule}, {Congiu}, {Dulieu},
  {Bonnin}, {Bassas}, {Duvernay}, {Danger}, \& {Chiavassa}}]{Noble2015}
{Noble}, J.~A., {Theule}, P., {Congiu}, E., {et~al.} 2015, \aap, 576, A91

\bibitem[{{Parise} {et~al.}(2006){Parise}, {Ceccarelli}, {Tielens}, {Castets},
  {Caux}, {Lefloch}, \& {Maret}}]{Parise2006}
{Parise}, B., {Ceccarelli}, C., {Tielens}, A.~G.~G.~M., {et~al.} 2006, \aap,
  453, 949

\bibitem[{{Pickett} {et~al.}(1998){Pickett}, {Poynter}, {Cohen}, {Delitsky},
  {Pearson}, \& {M{\"u}ller}}]{Pickett1998}
{Pickett}, H.~M., {Poynter}, R.~L., {Cohen}, E.~A., {et~al.} 1998, \jqsrt, 60,
  883

\bibitem[{{Raunier} {et~al.}(2004){Raunier}, {Chiavassa}, {Duvernay}, {Borget},
  {Aycard}, {Dartois}, \& {d'Hendecourt}}]{Raunier2004}
{Raunier}, S., {Chiavassa}, T., {Duvernay}, F., {et~al.} 2004, \aap, 416, 165

\bibitem[{{Redondo} {et~al.}(2014{\natexlab{a}}){Redondo}, {Barrientos}, \&
  {Largo}}]{Redondo2014a}
{Redondo}, P., {Barrientos}, C., \& {Largo}, A. 2014{\natexlab{a}}, \apj, 793,
  32

\bibitem[{{Redondo} {et~al.}(2014{\natexlab{b}}){Redondo}, {Barrientos}, \&
  {Largo}}]{Redondo2014b}
{Redondo}, P., {Barrientos}, C., \& {Largo}, A. 2014{\natexlab{b}}, \apj, 780,
  181

\bibitem[{{Richard} {et~al.}(2013){Richard}, {Margul{\`e}s}, {Caux}, {Kahane},
  {Ceccarelli}, {Guillemin}, {Motiyenko}, {Vastel}, \& {Groner}}]{Richard2013}
{Richard}, C., {Margul{\`e}s}, L., {Caux}, E., {et~al.} 2013, \aap, 552, A117

\bibitem[{{Saladino} {et~al.}(2012){Saladino}, {Crestini}, {Pino}, {Costanzo},
  \& {Di Mauro}}]{Saladino2012}
{Saladino}, R., {Crestini}, C., {Pino}, S., {Costanzo}, G., \& {Di Mauro}, E.
  2012, Physics of Life Reviews, 9, 84

\bibitem[{{Vorob'eva} \& {Dyubko}(1994)}]{Vorobeva1994}
{Vorob'eva}, E.~M. \& {Dyubko}, S.~F. 1994, Radiophysics and Quantum
  Electronics, 37, 155

\bibitem[{{Wilson}(1999)}]{Wilson1999}
{Wilson}, T.~L. 1999, Reports on Progress in Physics, 62, 143

\bibitem[{{Yamaguchi} {et~al.}(2012){Yamaguchi}, {Takano}, {Watanabe}, {Sakai},
  {Sakai}, {Liu}, {Su}, {Hirano}, {Takakuwa}, {Aikawa}, {Nomura}, \&
  {Yamamoto}}]{Yamaguchi2012}
{Yamaguchi}, T., {Takano}, S., {Watanabe}, Y., {et~al.} 2012, \pasj, 64, 105

\end{thebibliography}
\newpage
\appendix
%



\begin{appendix}

\section{Spectroscopic data}

A list of unblended and optically thin lines used in the analysis is presented in Table \ref{table_obs}. The spectroscopic data for NH$_2$CHO $\varv$=0, NH$_2$CHO $\varv_{12}$=1, NH$_2$$^{13}$CHO, $^{15}$NH$_2$CHO, NH$_2$CH$^{18}$O, NH$_2$CDO, cis-NHDCHO, trans-NHDCHO \citep{Kurland1957,Kukolich1971,Hirota1974,Gardner1980,Moskienko1991,Vorobeva1994,Blanco2006,Kryvda2009,Motiyenko2012,Kutsenko2013} and HNCO \citep{Kukolich1971,Hocking1975,Niedenhoff1995,Lapinov2007} come from the Cologne Database for Molecular Spectroscopy \citep[CMDS,][]{Muller2001,Muller2005}, while the data for DNCO, HN$^{13}$CO, H$^{15}$NCO and HNC$^{18}$O \citep{Hocking1975} are taken from the Jet Propulsion Laboratory (JPL) database \citep{Pickett1998}.  
It should be noted that there are significant differences for the predicted frequencies of the main isotopologue of NH$_2$CHO between CDMS and JPL ($>$ 1 MHz).  
A better agreement is found with the observations for the most recent entry in CDMS.
For some of the HNCO isotopologues, there is a lack of published spectroscopic data at high frequencies. In particular for H$^{15}$NCO, the range of uncertainty for some of the frequencies is quite high.
As the HN$^{13}$CO transitions all appeared  slightly shifted, compared to the observations, we applied a correction of $+0.5$ MHz to model the lines. 

The column densities of the formamide isotopologues given in Table \ref{results_model} were corrected by a factor of 1.5 to take into account the contribution of the vibrational states for an excitation temperature of 300 K.

\onecolumn 
\begin{longtab} 
\begin{longtable}{lccccccc}
\caption{\label{table_obs} Detected lines of NH$_2$CHO, HNCO and their isotopologues used in the analysis\tablefootmark{(a)}.} \\
\hline \hline
Species & Transition & Frequency & $E_{\rm up}$ & $A_{\rm ij}$ & $g_{\rm up}$ & \\
 & & (MHz) & (K) & (s$^{-1}$) & & \\
 \hline
 \endfirsthead
\caption{continued.} \\
 \hline\hline
 Species & Transition & Frequency & $E_{\rm up}$ & $A_{\rm ij}$ & $g_{\rm up}$ & \\
 & & (MHz) & (K) & (s$^{-1}$) & & \\
 \hline 
 \endhead
 \hline
 \endfoot
NH$_2$CDO & (17 0 17 -- 16 0 16) & 329995.2 & 145.6 & 2.64\,$\times$\,10$^{-3}$ & 105 & \\ 
NH$_2$CDO & (16 9 7 -- 15 9 6) & 333363.6 & 308.9 & 1.87\,$\times$\,10$^{-3}$ & 99 & \\ 
NH$_2$CDO & (16 9 8 -- 15 9 7) & 333363.6 & 308.9 & 1.87\,$\times$\,10$^{-3}$ & 99 &  \\
NH$_2$CDO & (16 7 10 -- 15 7 9) & 333696.6 & 240.7 & 2.22\,$\times$\,10$^{-3}$ & 99 & \\ 
NH$_2$CDO & (16 7 9 -- 15 7 8) & 333696.6 & 240.7 & 2.22\,$\times$\,10$^{-3}$ & 99 & \\
NH$_2$CDO & (16 4 13 -- 15 4 12) & 335234.9 & 170.5 & 2.61\,$\times$\,10$^{-3}$ & 99 & \\
NH$_2$CDO & (16 3 13 -- 15 3 12) & 342320.7 & 156.9 & 2.86\,$\times$\,10$^{-3}$ & 99 & \\
NH$_2$CDO & (17 1 16 -- 16 1 15) & 351988.3 & 158.1 & 3.18\,$\times$\,10$^{-3}$ & 105 & \\
NH$_2$CDO & (17 10 7 -- 16 10 6) & 354151.5 & 366.4 & 2.15\,$\times$\,10$^{-3}$ & 105 & \\
NH$_2$CDO & (17 10 8 -- 16 10 7) & 354151.5 & 366.4 & 2.15\,$\times$\,10$^{-3}$ & 105 &  \\
NH$_2$CDO & (17 9 8 -- 16 9 7) & 354257.0 & 325.9 & 2.37\,$\times$\,10$^{-3}$ & 105 & \\ 
NH$_2$CDO & (17 9 9 -- 16 9 8) & 354257.0 & 325.9 & 2.37\,$\times$\,10$^{-3}$ & 105 &  \\
NH$_2$CDO & (17 8 10 -- 16 8 9) & 354416.0 & 289.6 & 2.56\,$\times$\,10$^{-3}$ & 105 & \\ 
NH$_2$CDO & (17 8 9 -- 16 8 8) & 354416.0 & 289.6 & 2.56\,$\times$\,10$^{-3}$ & 105 &  \\
NH$_2$CDO & (17 7 11 -- 16 7 10) & 354661.3 & 257.7 & 2.74\,$\times$\,10$^{-3}$ & 105 & \\
NH$_2$CDO & (17 7 10 -- 16 7 9) & 354661.3 & 257.7 & 2.74\,$\times$\,10$^{-3}$ & 105 &  \\
NH$_2$CDO & (17 5 12 -- 16 5 11) & 355800.2 & 206.7 & 3.04\,$\times$\,10$^{-3}$ & 105 & \\
NH$_2$CDO & (17 4 13 -- 16 4 12) & 357938.5 & 187.8 & 3.20\,$\times$\,10$^{-3}$ & 105 & \\
\hline
cis-NHDCHO & (16 3 13 -- 15 3 12) & 331372.8 & 156.0 & 2.59\,$\times$\,10$^{-3}$ & 99 & \\
cis-NHDCHO & (16 2 14 -- 15 2 13) & 337248.5 & 146.0 & 2.79\,$\times$\,10$^{-3}$ & 99 & \\
cis-NHDCHO & (17 2 16 -- 16 2 15) & 340520.3 & 158.0 & 2.87\,$\times$\,10$^{-3}$ & 105 & \\
cis-NHDCHO & (18 1 18 -- 17 1 17) & 344878.9 & 160.8 & 3.02\,$\times$\,10$^{-3}$ & 111 & \\
cis-NHDCHO & (17 8 10 -- 16 8 9) & 346444.0 & 306.6 & 2.39\,$\times$\,10$^{-3}$ & 105 & \\
cis-NHDCHO & (17 8 9 -- 16 8 8) & 346444.0 &  306.6 & 2.39\,$\times$\,10$^{-3}$ & 105 & \\
cis-NHDCHO & (17 7 11 -- 16 7 10) & 346586.8 & 269.8 & 2.56\,$\times$\,10$^{-3}$ & 105 & \\
cis-NHDCHO & (17 7 10 -- 16 7 9) & 346586.8 &  269.8 & 2.56\,$\times$\,10$^{-3}$ & 105 &\\
cis-NHDCHO & (17 6 12 -- 16 6 11) & 346826.8 & 238.0 & 2.70\,$\times$\,10$^{-3}$ & 105 & \\
cis-NHDCHO & (17 6 11 -- 16 6 10) & 346827.5 & 238.0 & 2.70\,$\times$\,10$^{-3}$ & 105 &  \\
cis-NHDCHO & (17 3 15 -- 16 3 14) & 347115.8 & 172.0 & 2.99\,$\times$\,10$^{-3}$ & 105 & \\
cis-NHDCHO & (17 5 12 -- 16 5 11) & 347268.9 & 211.1 & 2.83\,$\times$\,10$^{-3}$ & 105 & \\
cis-NHDCHO & (17 4 14 -- 16 4 13) & 347827.8 & 189.2 & 2.94\,$\times$\,10$^{-3}$ & 105 & \\
cis-NHDCHO & (17 3 14 -- 16 3 13) & 353047.5 & 173.0 & 3.15\,$\times$\,10$^{-3}$ & 105 & \\
\hline
trans-NHDCHO & (17 8 9 -- 16 8 8) & 333628.6 & 332.4 & 2.14\,$\times$\,10$^{-3}$ & 105 & \\
trans-NHDCHO & (17 8 10 -- 16 8 9) & 333628.6 &  332.4 & 2.14\,$\times$\,10$^{-3}$ & 105 & \\
trans-NHDCHO & (17 7 11 -- 16 7 10) & 333694.1 & 288.3 & 2.28\,$\times$\,10$^{-3}$ & 105 & \\
trans-NHDCHO & (17 7 10 -- 16 7 9) & 333694.1 & 288.3 & 2.28\,$\times$\,10$^{-3}$ & 105 & \\
trans-NHDCHO & (17 6 12 -- 16 6 11) & 333812.6 & 250.1 & 2.41\,$\times$\,10$^{-3}$ & 105 & \\
trans-NHDCHO & (17 6 11 -- 16 6 10) & 333812.7 &  250.1 & 2.41\,$\times$\,10$^{-3}$ & 105 & \\
trans-NHDCHO & (17 4 14 -- 16 4 13) & 334403.2 & 191.4 & 2.61\,$\times$\,10$^{-3}$ & 105 & \\
trans-NHDCHO & (18 1 18 -- 17 1 17) & 336945.3 & 157.3 & 2.82\,$\times$\,10$^{-3}$ & 111 & \\
trans-NHDCHO& (18 0 18 -- 17 0 17) & 338818.4 & 156.9 & 2.87\,$\times$\,10$^{-3}$ & 111 & \\
trans-NHDCHO & (17 1 16 -- 16 1 15) & 338878.8 & 150.6 & 2.86\,$\times$\,10$^{-3}$ & 105 & \\
trans-NHDCHO & (18 7 12 -- 17 7 11) & 353355.8 & 305.2 & 2.77\,$\times$\,10$^{-3}$ & 111 & \\
trans-NHDCHO & (18 7 11 -- 17 7 10) & 353355.8 &  305.2 & 2.77\,$\times$\,10$^{-3}$ & 111 & \\
trans-NHDCHO & (18 5 14 -- 17 5 13) & 353758.4 & 234.7 & 3.02\,$\times$\,10$^{-3}$ & 111 & \\
trans-NHDCHO & (18 3 16 -- 17 3 15) & 354028.8 & 187.8 & 3.19\,$\times$\,10$^{-3}$ & 111 & \\
trans-NHDCHO & (18 4 15 -- 17 4 14) & 354185.9 & 208.4 & 3.13\,$\times$\,10$^{-3}$ & 111 & \\
 \hline 
 NH$_2$$^{13}$CHO & (16 10 6 -- 15 10 5) & 339170.1 & 427.9 & 1.75\,$\times$\,10$^{-3}$ & 33 & \\
NH$_2$$^{13}$CHO & (16 10 7 -- 15 10 6) & 339170.1 & 427.9 & 1.75\,$\times$\,10$^{-3}$ & 33 & \\
NH$_2$$^{13}$CHO & (16 9 7 -- 15 9 6) & 339179.6 & 373.0 & 1.97\,$\times$\,10$^{-3}$ & 33 & \\
NH$_2$$^{13}$CHO & (16 9 8 -- 15 9 7) & 339179.6 & 373.0 & 1.97\,$\times$\,10$^{-3}$ & 33 &  \\
NH$_2$$^{13}$CHO & (16 8 8 -- 15 8 7) & 339213.5 & 323.8 & 2.16\,$\times$\,10$^{-3}$ & 33 & \\
NH$_2$$^{13}$CHO & (16 8 9 -- 15 8 8) & 339213.5 & 323.8 & 2.16\,$\times$\,10$^{-3}$ & 33 &  \\
NH$_2$$^{13}$CHO & (16 5 11 -- 15 5 10) & 339672.1 & 210.9 & 2.61\,$\times$\,10$^{-3}$ & 33 & \\
NH$_2$$^{13}$CHO & (16 4 13 -- 15 4 12) & 340090.4 & 184.9 & 2.72\,$\times$\,10$^{-3}$ & 33 & \\
NH$_2$$^{13}$CHO & (16 4 12 -- 15 4 11) & 340273.4  & 184.9 & 2.73\,$\times$\,10$^{-3}$ & 33 & \\
NH$_2$$^{13}$CHO & (17 1 17 -- 16 1 16) & 342156.0 & 151.5 & 2.95\,$\times$\,10$^{-3}$ & 35 & \\
NH$_2$$^{13}$CHO & (17 9 8 -- 16 9 7) & 360396.3 & 390.3 & 2.49\,$\times$\,10$^{-3}$ & 35 & \\
NH$_2$$^{13}$CHO & (17 9 9 -- 16 9 8) & 360396.3 & 390.3 & 2.49\,$\times$\,10$^{-3}$ & 35 &  \\
NH$_2$$^{13}$CHO & (17 7 11 -- 16 7 10) & 360531.8 & 297.7 & 2.88\,$\times$\,10$^{-3}$ & 35 & \\
NH$_2$$^{13}$CHO & (17 7 10 -- 16 7 9) & 360531.8 & 297.7 & 2.88\,$\times$\,10$^{-3}$ & 35 &  \\
NH$_2$$^{13}$CHO & (18 1 18 -- 17 1 17) & 361904.8 & 168.9 & 3.49\,$\times$\,10$^{-3}$ & 37 & \\
 \hline
NH$_2$CHO $\varv$=0 & (16 3 14 -- 16 2 15) & 331685.9 & 165.6 & 7.87\,$\times$\,10$^{-5}$ & 33 & \\
NH$_2$CHO $\varv$=0 & (8 2 7 -- 7 1 6) & 334483.5 & 48.5 & 5.49\,$\times$\,10$^{-5}$ & 17 & \\
NH$_2$CHO $\varv$=0 & (17 3 15 -- 17 2 16) & 336733.0 & 183.0 & 8.2\,$\times$\,10$^{-5}$ & 35 & \\
NH$_2$CHO $\varv$=0 & (34 3 31 -- 34 2 32) & 342029.5 & 645.9 & 1.07\,$\times$\,10$^{-4}$ & 69 & \\
NH$_2$CHO $\varv$=0 & (18 3 16 -- 18 2 17) & 342511.1 & 201.3 & 8.57\,$\times$\,10$^{-5}$ & 37 & \\
NH$_2$CHO $\varv$=0 & (28 4 24 -- 28 3 25) & 344545.8 & 464.1 & 1.15\,$\times$\,10$^{-4}$ & 57 & \\
NH$_2$CHO $\varv$=0 & (19 3 17 -- 19 2 18) & 349051.7 & 220.7 & 8.99\,$\times$\,10$^{-5}$ & 39 & \\
NH$_2$CHO $\varv$=0 & (20 3 18 -- 20 2 19) & 356379.8 & 241.1 & 9.47\,$\times$\,10$^{-5}$ & 41 & \\
NH$_2$CHO $\varv$=0 & (20 1 19 -- 19 2 18) & 359119.4 & 221.2 & 8.45\,$\times$\,10$^{-5}$ & 41 & \\
\hline
NH$_2$CHO $\varv_{12}$=1 & (17 14 3 -- 16 14 2) & 360717.7 & 1144.3 & 1.12\,$\times$\,10$^{-3}$ & 35 & \\
NH$_2$CHO $\varv_{12}$=1 & (17 14 4 -- 16 14 3) & 360717.7 & 1144.3 & 1.12\,$\times$\,10$^{-3}$ & 35 & \\
 \hline 
DNCO & (17 1 17 18 -- 16 1 16 17) & 344629.4 & 172.9 & 5.92\,$\times$\,10$^{-4}$ & 37 & \\
DNCO & (17 1 17 17 -- 16 1 16 16) & 344629.4 & 172.9 & 5.90\,$\times$\,10$^{-4}$ & 35 &  \\
DNCO & (17 1 17 16 -- 16 1 16 15) & 344629.4 & 172.9 & 5.90\,$\times$\,10$^{-4}$ & 33 &  \\
DNCO & (17 0 17 18 -- 16 0 16 17) & 346556.2 & 149.7 & 6.04\,$\times$\,10$^{-4}$ & 37 & \\
DNCO & (17 0 17 17 -- 16 0 16 16) & 346556.2 & 149.7 & 6.02\,$\times$\,10$^{-4}$ & 35 &  \\
DNCO & (17 0 17 16 -- 16 0 16 15) & 346556.2 & 149.7 & 6.02\,$\times$\,10$^{-4}$ & 33 &  \\
DNCO & (17 5 12 18 -- 16 5 11 17) & 346714.9 & 750.6 & 5.53\,$\times$\,10$^{-4}$ & 37 & \\
DNCO & (17 5 13 18 -- 16 5 12 17) & 346714.9 & 750.6 & 5.53\,$\times$\,10$^{-4}$ & 37 &  \\
DNCO & (17 5 13 16 -- 16 5 12 15) & 346714.9 & 750.6 & 5.50\,$\times$\,10$^{-4}$ & 33 &  \\
DNCO & (17 5 12 16 -- 16 5 11 15) & 346714.9 & 750.6 & 5.50\,$\times$\,10$^{-4}$ & 33 &  \\
DNCO & (17 5 13 17 -- 16 5 12 16) & 346714.9 & 750.6 & 5.51\,$\times$\,10$^{-4}$ & 35 &  \\
DNCO & (17 5 12 17 -- 16 5 11 16) & 346714.9 & 750.6 & 5.51\,$\times$\,10$^{-4}$ & 35 &  \\
DNCO & (17 1 16 18 -- 16 1 15 17) & 348599.7 & 174.6 & 6.13\,$\times$\,10$^{-4}$ & 37 & \\
DNCO & (17 1 16 17 -- 16 1 15 16) & 348599.7 & 174.6 & 6.10\,$\times$\,10$^{-4}$ & 35 &  \\
DNCO & (17 1 16 16 -- 16 1 15 15) & 348599.7 & 174.6 & 6.10\,$\times$\,10$^{-4}$ & 33 &  \\
\hline
HN$^{13}$CO & (15 2 13 16 -- 14 2 12 15) & 329594.5 & 299.2 & 5.08\,$\times$\,10$^{-4}$ & 33 & \\
HN$^{13}$CO & (15 2 13 14 -- 14 2 12 13) & 329594.5 & 299.2 & 5.06\,$\times$\,10$^{-4}$ & 29 &  \\
HN$^{13}$CO & (15 2 13 15 -- 14 2 12 14) & 329594.5 & 299.2 & 5.06\,$\times$\,10$^{-4}$ & 31 &  \\
HN$^{13}$CO & (15 0 15 16 -- 14 0 14 15) & 329673.4 & 126.6 & 5.18\,$\times$\,10$^{-4}$ & 33 & \\
HN$^{13}$CO & (15 0 15 15 -- 14 0 14 14) & 329673.4 & 126.6 & 5.16\,$\times$\,10$^{-4}$ & 31 &  \\
HN$^{13}$CO & (15 0 15 14 -- 14 0 14 13) & 329673.4 & 126.6 & 5.15\,$\times$\,10$^{-4}$ & 29 &  \\
HN$^{13}$CO & (15 1 14 16 -- 14 1 13 15) & 330860.2 & 170.2 & 5.21\,$\times$\,10$^{-4}$ & 33 & \\
HN$^{13}$CO & (15 1 14 14 -- 14 1 13 13) & 330860.2 & 170.2 & 5.19\,$\times$\,10$^{-4}$ & 29 &  \\
HN$^{13}$CO & (15 1 14 15 -- 14 1 13 14) & 330860.2 & 170.2 & 5.19\,$\times$\,10$^{-4}$ & 31 &  \\
HN$^{13}$CO & (16 1 16 17 -- 15 1 15 16) & 350340.3 & 186.1 & 6.20\,$\times$\,10$^{-4}$ & 35 & \\
HN$^{13}$CO & (16 1 16 16 -- 15 1 15 15) & 350340.3 & 186.1 & 6.18\,$\times$\,10$^{-4}$ & 33 &  \\
HN$^{13}$CO & (16 1 16 15 -- 15 1 15 14) & 350340.3 & 186.1 & 6.18\,$\times$\,10$^{-4}$ & 31 &  \\
HN$^{13}$CO & (16 3 14 17 -- 15 3 13 16) & 351427.6 & 531.9 & 6.07\,$\times$\,10$^{-4}$ & 35 & \\
HN$^{13}$CO & (16 3 14 15 -- 15 3 13 14) & 351427.6 & 531.9 & 6.04\,$\times$\,10$^{-4}$ & 31 &  &  \\
HN$^{13}$CO & (16 3 14 16 -- 15 3 13 15) & 351427.7 & 531.9 & 6.04\,$\times$\,10$^{-4}$ & 33 &  &  \\
HN$^{13}$CO & (16 3 13 17 -- 15 3 12 16) & 351427.7 & 531.9 & 6.07\,$\times$\,10$^{-4}$ & 35 &  &  \\
HN$^{13}$CO & (16 3 13 15 -- 15 3 12 14) & 351427.7 & 531.9 & 6.04\,$\times$\,10$^{-4}$ & 31 &  &  \\
HN$^{13}$CO & (16 3 13 16 -- 15 3 12 15) & 351427.7 & 531.9 & 6.04\,$\times$\,10$^{-4}$ & 33 &  &  \\
HN$^{13}$CO & (16 2 15 17 -- 15 2 14 16) & 351548.3 & 316.1 & 6.19\,$\times$\,10$^{-4}$ & 35 & \\
HN$^{13}$CO & (16 2 15 15 -- 15 2 14 14) & 351548.3 & 316.1 & 6.17\,$\times$\,10$^{-4}$ & 31 &  &  \\
HN$^{13}$CO & (16 2 15 16 -- 15 2 14 15) & 351548.3 & 316.1 & 6.17\,$\times$\,10$^{-4}$ & 33 &  &  \\
HN$^{13}$CO & (16 2 14 17 -- 15 2 13 16) & 351561.8 & 316.1 & 6.19\,$\times$\,10$^{-4}$ & 35 & \\
HN$^{13}$CO & (16 2 14 15 -- 15 2 13 14) & 351561.8 & 316.1 & 6.17\,$\times$\,10$^{-4}$ & 31 &  &   \\
HN$^{13}$CO & (16 2 14 16 -- 15 2 13 15) & 351561.8 & 316.1 & 6.17\,$\times$\,10$^{-4}$ & 33 &  &  \\
HN$^{13}$CO & (16 2 14 17 -- 15 2 13 16) & 351561.8 & 316.1 & 6.19\,$\times$\,10$^{-4}$ & 35 &  &  \\
HN$^{13}$CO & (16 2 14 15 -- 15 2 13 14) & 351561.8 & 316.1 & 6.17\,$\times$\,10$^{-4}$ & 31 &  \\
HN$^{13}$CO & (16 2 14 16 -- 15 2 13 15) & 351561.8 & 316.1 & 6.17\,$\times$\,10$^{-4}$ & 33 &  \\
HN$^{13}$CO & (16 0 16 17 -- 15 0 15 16) & 351642.9 & 143.5 & 6.30\,$\times$\,10$^{-4}$ & 35 & \\
HN$^{13}$CO & (16 0 16 16 -- 15 0 15 15) & 351642.9 & 143.5 & 6.27\,$\times$\,10$^{-4}$ & 33 &  \\
HN$^{13}$CO & (16 0 16 15 -- 15 0 15 14) & 351642.9 & 143.5 & 6.27\,$\times$\,10$^{-4}$ & 31 &  \\
\hline
\end{longtable}
\tablefoot{
\tablefoottext{a}{This list only includes optically thin and unblended lines.}}
\end{longtab}

\section{Additional figures}

\begin{figure*}[h!]
\includegraphics[width=0.95\textwidth]{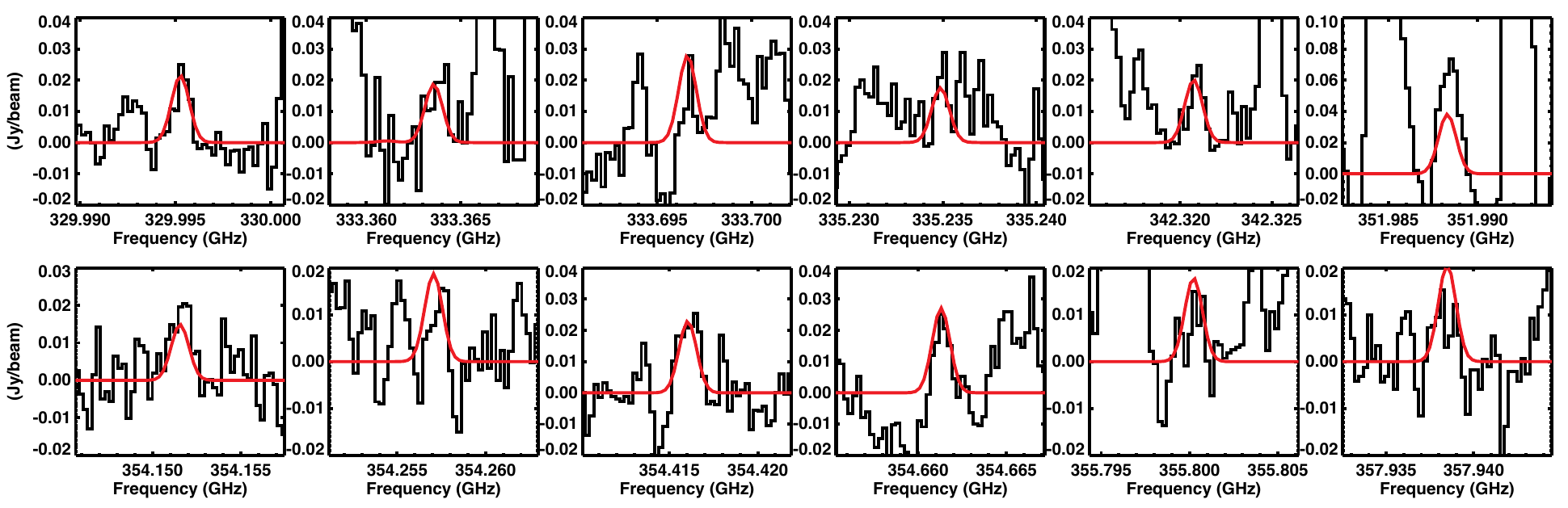}
\caption{{\it Black}: Detected lines of NH$_2$CDO. {\it Red}: Best-fit model for $T_{\rm ex}$=300\,K.}
\label{fig:NH2CDO}
\end{figure*}

\begin{figure*}[h!]
\includegraphics[width=0.95\textwidth]{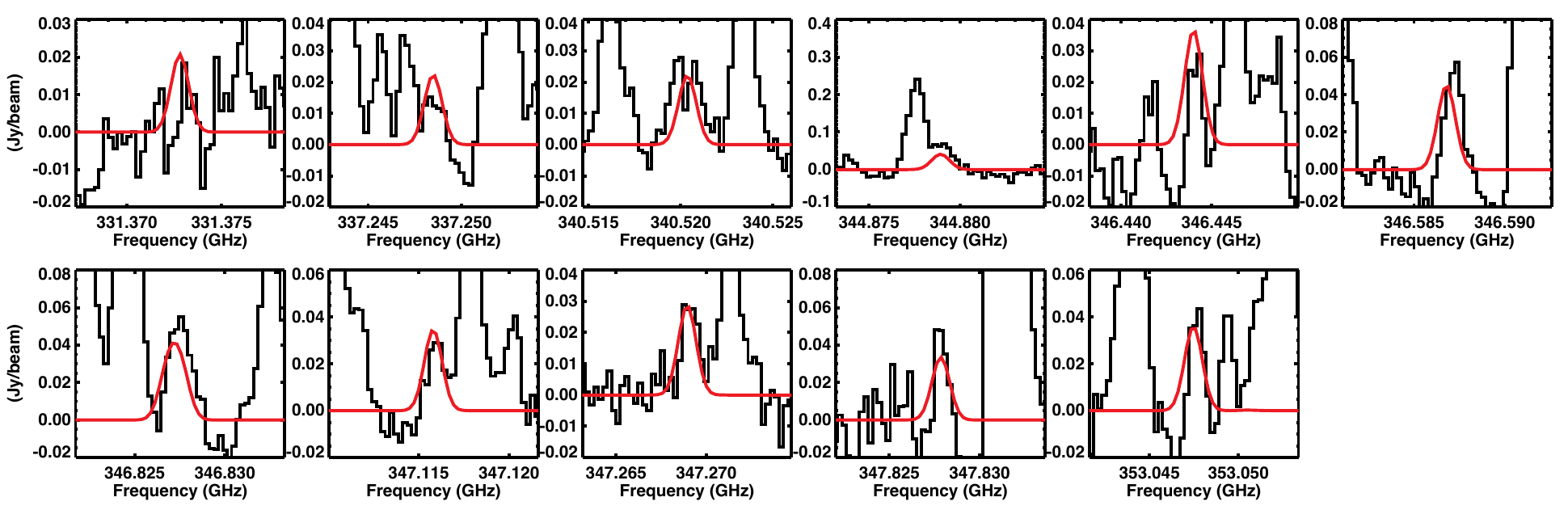}
\caption{{\it Black}: Detected lines of cis-NHDCHO. {\it Red}: Best-fit model for $T_{\rm ex}$=300\,K.}
\label{fig:cis-NHDCHO}
\end{figure*}

\begin{figure*}[h!]
\includegraphics[width=0.95\textwidth]{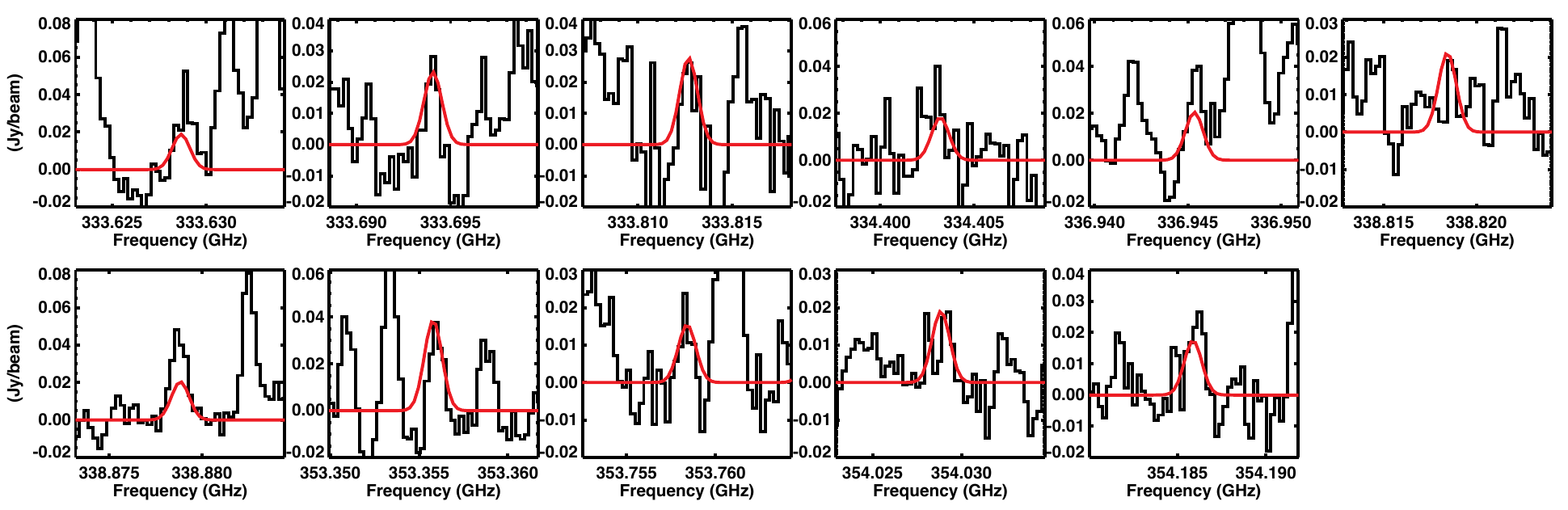}
\caption{{\it Black}: Detected lines of trans-NHDCHO. {\it Red}: Best-fit model for $T_{\rm ex}$=300\,K.}
\label{fig:trans-NHDCHO}
\end{figure*}

\begin{figure*}[h!]
\includegraphics[width=0.95\textwidth]{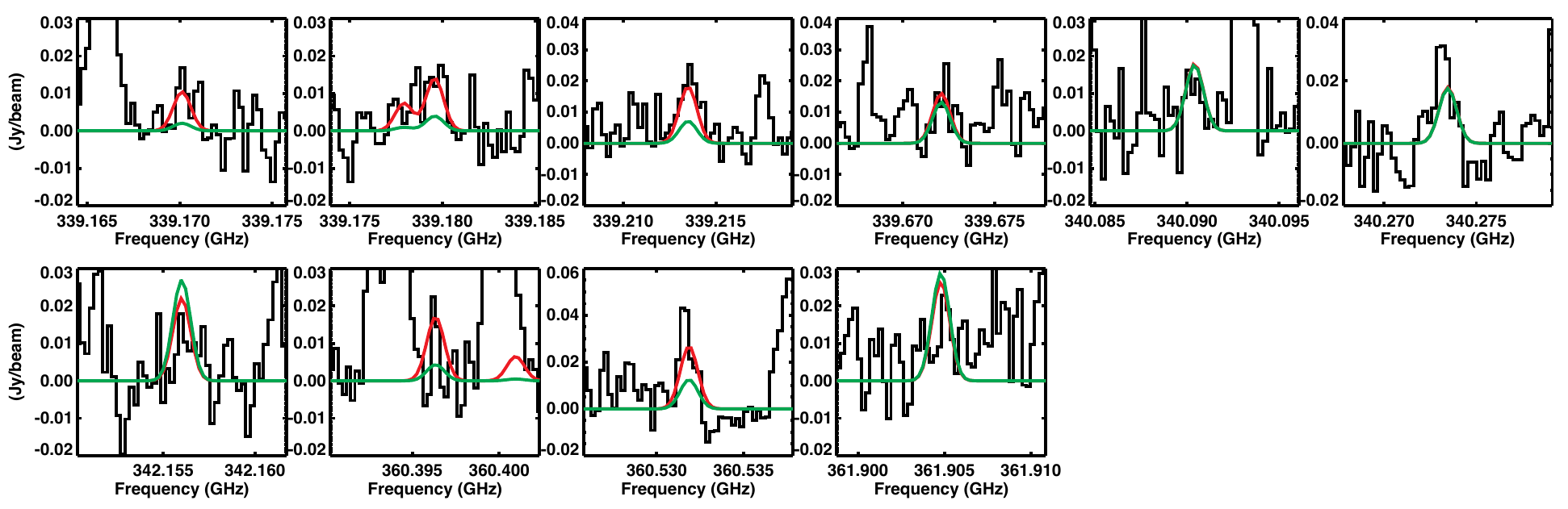}
\caption{{\it Black}: Detected lines of NH$_2$$^{13}$CHO. {\it Red}: Best-fit model for $T_{\rm ex}$=300\,K. {\it Green}: Best-fit model for $T_{\rm ex}$=100\,K.}
\label{fig:NH213CHO}
\end{figure*}

\begin{figure*}[h!]
\includegraphics[width=0.95\textwidth]{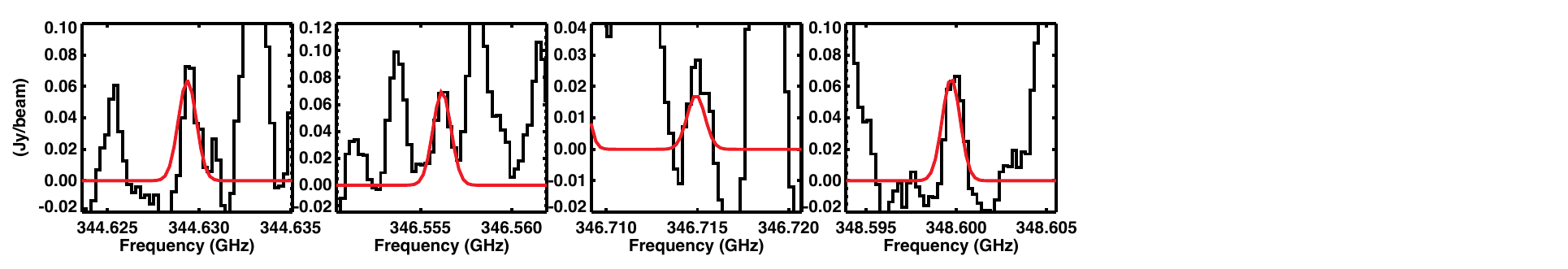}
\caption{{\it Black}: Detected lines of DNCO. {\it Red}: Best-fit model for $T_{\rm ex}$=300\,K.}
\label{fig:DNCO}
\end{figure*}

\begin{figure*}[h!]
\includegraphics[width=0.95\textwidth]{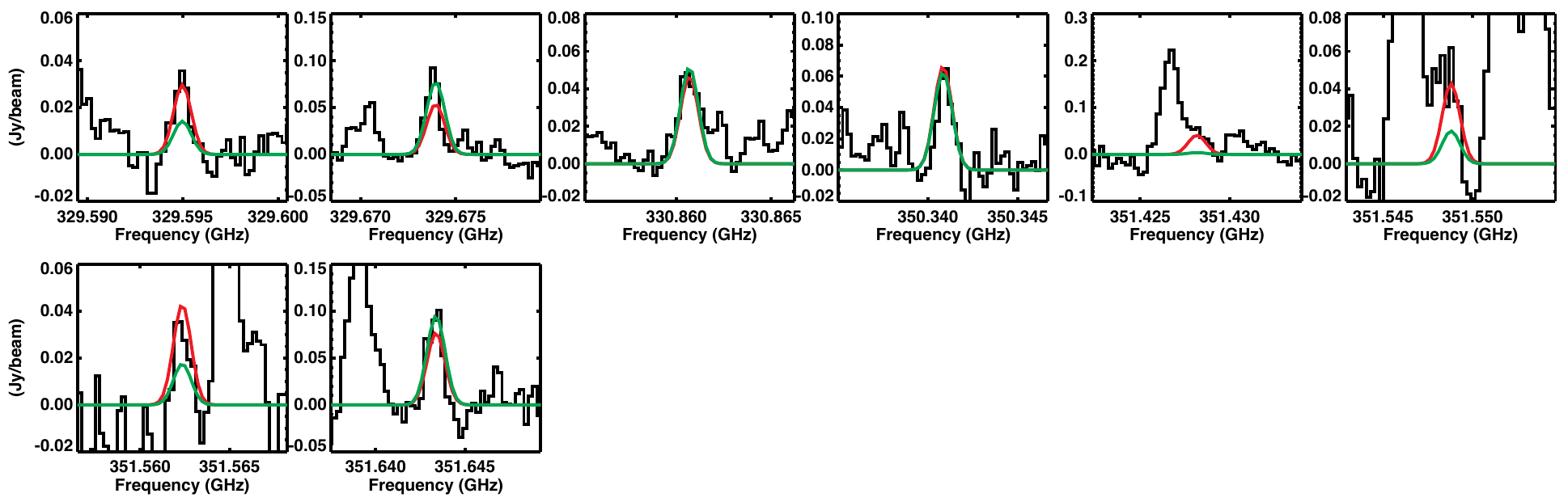}
\caption{{\it Black}: Detected lines of HN$^{13}$CO. {\it Red}: Best-fit model for $T_{\rm ex}$=300\,K. {\it Green}: Best-fit model for $T_{\rm ex}$=100\,K.}
\label{fig:HN13CO}
\end{figure*}

\begin{figure*}[h!]
\includegraphics[width=0.95\textwidth]{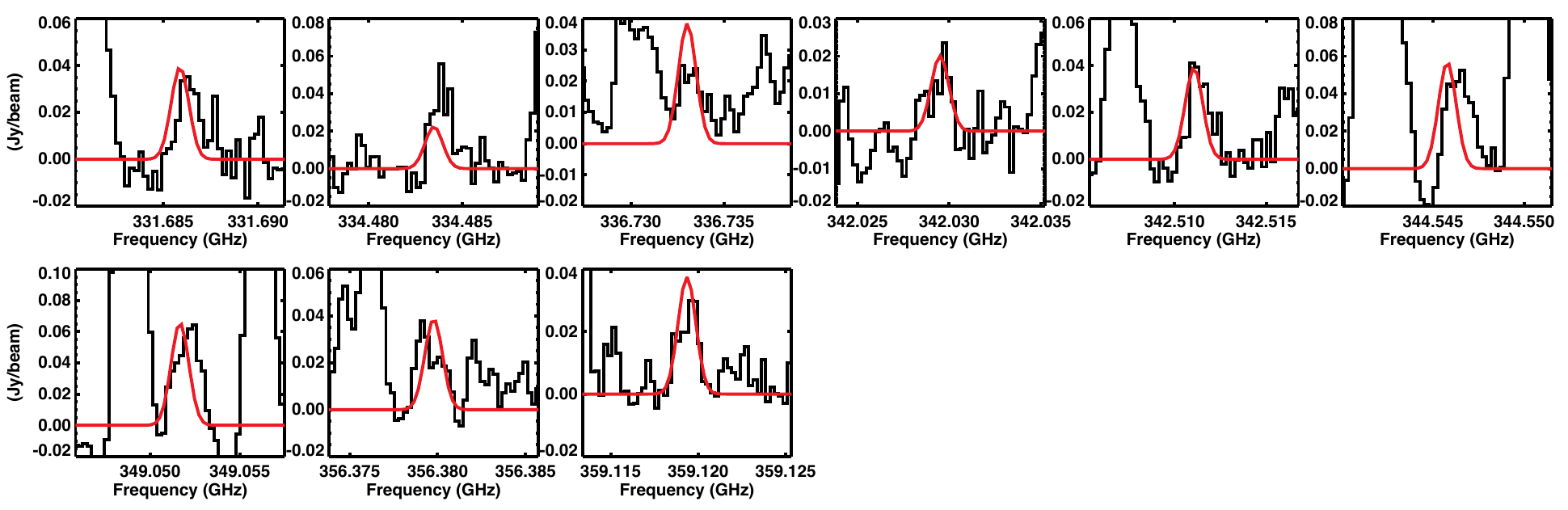}
\caption{{\it Black}: Lines of NH$_2$CHO $\varv$=0 with the lowest opacities. {\it Red}: Model based on the analysis of the NH$_2$$^{13}$CHO lines and a $^{12}$C/$^{13}$C ratio equal to 68.}
\label{fig:HCONH2_v0}
\end{figure*}

\begin{figure*}[h!]
\includegraphics[width=0.95\textwidth]{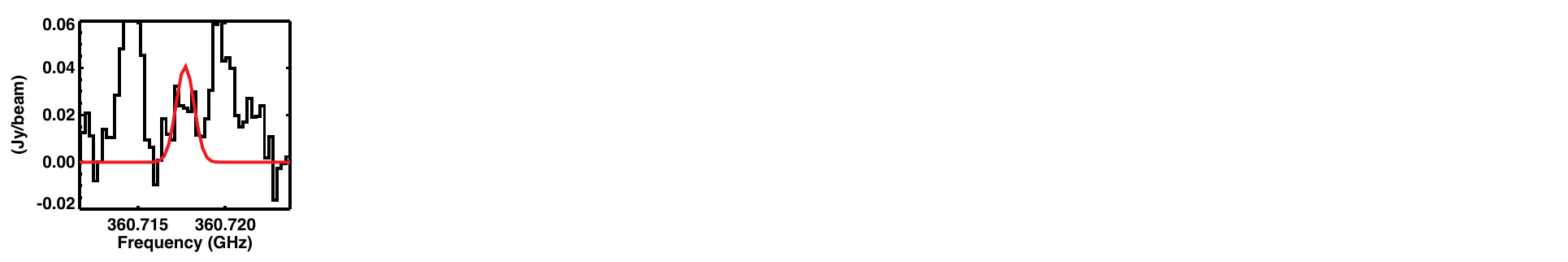}
\caption{{\it Black}: Line of NH$_2$CHO $\varv_{12}$=1 with the lowest opacity. {\it Red}: Model based on the analysis of the NH$_2$$^{13}$CHO lines and a $^{12}$C/$^{13}$C ratio equal to 68.}
\label{fig:HCONH2_v12}
\end{figure*}


\end{appendix}

\end{document}